# DIFFERENT NETWORK TOPOLOGIES FOR DISTRIBUTED ELECTRIC DAMPING OF BEAM VIBRATIONS.

M.Porfiri[1], C.Maurini[2], F.dell'Isola[1], J. Pouget[2]

[1] *Dipartimento di Ingegneria Strutturale & Geotecnica, Università di Roma "La Sapienza", via Eudossiana 18, 00184 Roma, Italy*

[2] *Laboratoire d'Etudes Mécanique des Assemblages (FRE 2481), Université de Versailles/Saint-Quentin-en Yvelines, 45, ave. des Etats-Unis, 78035 Versailles cedex, France*

*Summary* In this work passive electric damping of structural vibrations by distributed piezoelectric transducers and electric networks is analyzed. Different distributed electric controllers are examined as finite degrees of freedom systems and their performances are compared. Modal reduction is used to optimize the electric parameters.

Passive vibration control can be achieved by coupling a mechanical structure S to an auxiliary dissipative system S' by means of an appropriate transduction device T. Once the properties of T are given, a proper design of the auxiliary system S' allows to enhance the energy exchange between S and S' and the energy dissipation in S' by exploiting an internal resonance phenomenon in (S,T,S'). In the technical literature the additional system S' is called dynamic vibration absorber. For example, the flexural vibrations of an Euler beam can be controlled connecting to it a proper spring-mass-damper system. The same effect can be achieved by bonding on the beam a piezoelectric transducer shunted with a resistor and an inductor, realizing the so called resonant shunted piezoelectric transducer (see Hagood and Von Flotow [1]). Indeed, the resistor and the inductor together with the capacitance of the piezoelectric transducer form a RLC circuit piezoelectrically coupled to the beam oscillations.

An alternative approach suggests to couple the continuous system S to a distributed auxiliary system S' with distributed transduction devices. This idea has been exploited in F.dell'Isola and Vidoli [4] leading to the conception of a new smart structure, which is called piezo-electromechanical structure. It is conceived as a host structure on which an array of equally spaced piezoelectric transducers is positioned, and interconnected via an electrical network. In previous papers (Alessandroni et al. [3], Andreaus et al. [4], Maurini et al. [5]) the introduced systems are regarded as continuous electromechanical media and they are analyzed in the framework of homogenized models. This coarse model gives a synthetic description of the dynamical properties of the coupled systems and allows for a global characterization which has been used to deduce optimal interconnecting networks for fundamental structural elements (beams and plates). The validity of the results predicted by the homogenized model strongly depends on the number of used transducers and on the minimum wave-length of the considered dynamical phenomena.

In this work we will analyze the modular piezo-electromechanical systems from a different viewpoint. The electrical interconnecting network will be treated as a finite degrees of freedom system interacting with the continuous mechanical structure by means of a finite number of coupling elements. This strategy allows for a more accurate design of experimental prototypes, taking the problems related to the lumped nature of the electric system into account.

The methodologies described in this work will be immediately applied to a master problem: vibration damping of a cantilever beam. By assuming that a fixed number of piezoelectric elements is uniformly positioned on the host beam, different electrical interconnections will be designed, and their performances will be compared.

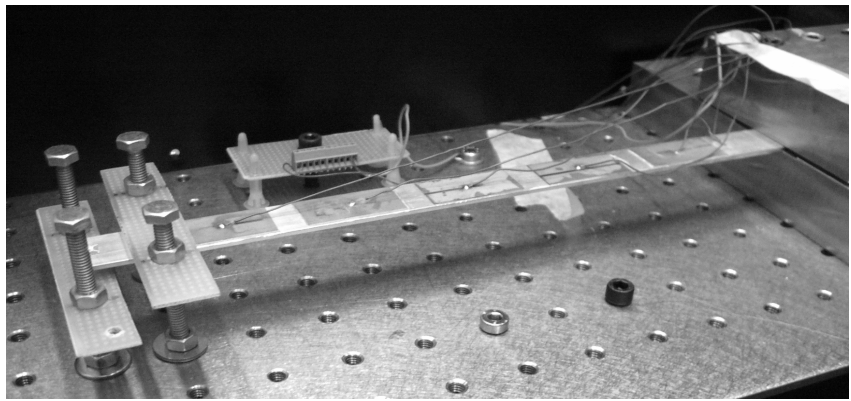

**Figure 1**. Beam with five uniformely distributed piezoelectric transducers.

The following representative electrical interconnections will be examined.
- *Single RL shunt*. All the piezoelectric elements are parallel connected and shunted with a single RL impedance: the auxiliary electric system S' is a one degree of freedom system.
- *Multiple RL shunt*. Each piezoelectric transducer is shunted with a separate RL impedance: the auxiliary system S' is a multi degrees of freedom system. However since no electrical interconnection is considered between the actuators, S' is a "diagonal" system.
- *Transmission line interconnection*. Each piezoelectric element is connected to the adjacent ones via a RL floating impedance: the auxiliary system S' is a multi degrees of freedom system. The electrical network including the piezoelectric capacitors represents a finite differences approximation of a classical transmission line. The electrical synergy of the transducers renders the presented configuration a basic example of a support for the propagation of electrical waves in a piezo-electromechanical system.

The analysis of the sketched configurations will include the following steps:
a) modal analysis of the coupled electromechanical system and consistent modal reduction to a two DOF system;
b) optimization of the electrical parameters revisiting the design criteria of one DOF vibration absorbers;
c) validation of the modal reduction via numerical analyses on a complete model.

## References

[1] N. W. Hagood and A.von Flotow, Damping of structural vibrations with piezoelectric materials and passive electrical networks, *Journal of Sound and Vibrations* **146** (1991) 243-268.

[2] F.dell'Isola and S.Vidoli, Continuum modelling of piezoelectromechanical truss beams: an application to vibration damping, *Archive of Applied Mechanics* **68** (1998) 1-19.

[3] S.Alessandroni, F.dell'Isola, M.Porfiri, A revival of electric analogs for vibrating mechanical systems aimed to their efficient control by PZT actuators, *International Journal of Solids and Structures* **39** (20) (2002) 5295-5324.

[4] U.Andreaus, F.dell'Isola, M.Porfiri, Piezoelectric passive distributed controllers for beam flexural vibrations, to appear in *Journal of Vibration and Control*.

[5] C.Maurini, F.dell'Isola, D.Del Vescovo, Performance comparison of distributed beam vibration absorbers synthesized by piezoelectric transducers and electr(on)ic networks, *Mechanical Systems and Signal Processing*, in press.